\documentclass[journal,twocolumn,10pt]{IEEEtran}
\IEEEoverridecommandlockouts
\usepackage{amssymb}
\usepackage{setspace}
\usepackage{amsmath}
\usepackage{graphicx}
\usepackage{color} 
\usepackage{transparent}
\usepackage{float}
\usepackage{subfigure}
\usepackage{afterpage}

\usepackage{epstopdf}
\usepackage{stfloats}
\usepackage{url}
\usepackage[T1]{fontenc}
\usepackage{multirow}
\usepackage{hhline}
\usepackage{booktabs}
\usepackage{color}

\usepackage{etoolbox}

\makeatother

\ifCLASSINFOpdf
  
\else
  
\fi
\begin{document}
\title{Adaptive Nonlinear RF Cancellation for Improved Isolation in Simultaneous Transmit-Receive Systems}
\author{Adnan~Kiayani,~\IEEEmembership{Member,~IEEE,}
        Muhammad~Zeeshan~Waheed,
				Lauri~Anttila,~\IEEEmembership{Member,~IEEE,}
        Mahmoud~Abdelaziz,~\IEEEmembership{Student Member,~IEEE,}
				Dani~Korpi,~\IEEEmembership{Student Member,~IEEE,}
				Ville~Syrj\"{a}l\"{a},~\IEEEmembership{Member,~IEEE,}
				~Marko~Kosunen,~\IEEEmembership{Member,~IEEE,}
				~Kari~Stadius,~\IEEEmembership{Member,~IEEE,}
				~Jussi~Ryyn\"{a}nen,~\IEEEmembership{Member,~IEEE,}
				and~Mikko~Valkama,~\IEEEmembership{Senior Member,~IEEE}
\thanks{This work was supported by the Academy of Finland (under the projects 304147 ``In-Band Full-Duplex Radio Technology: Realizing Next Generation Wireless Transmission'', and 301820 ``Competitive Funding to Strengthen University Research Profiles''), the Finnish Funding Agency for Innovation (Tekes, under the project ``5G Transceivers for Base Stations and Mobile Devices (5G TRx)''), Nokia Bell Labs, TDK-EPCOS, Pulse, Sasken, and Huawei Technologies, Finland.}
\thanks{A. Kiayani, M. Waheed, L. Anttila, M. Abdelaziz, D. Korpi, V. Syrj\"{a}l\"{a}, and M. Valkama are with the Laboratory of Electronics and Communications Engineering, Tampere University of Technology, FI-33101 Tampere, Finland.} 
\thanks{M. Kosunen, K. Stadius, and J. Ryyn\"{a}nen are with the Department of Electronics and Nanoengineering, Aalto University, FI-00076 Espoo, Finland. (corresponding author e-mail: mikko.e.valkama@tut.fi).}} 

\maketitle
\begin{abstract}
This paper proposes an active radio frequency (RF) cancellation solution to suppress the transmitter (TX) passband leakage signal in radio transceivers supporting simultaneous transmission and reception. The proposed technique is based on creating an opposite-phase baseband equivalent replica of the TX leakage signal in the transceiver digital front-end through adaptive nonlinear filtering of the known transmit data, to facilitate highly accurate cancellation under a nonlinear power amplifier (PA). The active RF cancellation is then accomplished by employing an auxiliary transmitter chain, to generate the actual RF cancellation signal, and combining it with the received signal at the receiver (RX) low noise amplifier (LNA) input. A closed-loop parameter learning approach, based on the decorrelation learning rule, is also developed to efficiently estimate the coefficients of the nonlinear cancellation filter in the presence of a nonlinear PA with memory, finite passive isolation, and a nonlinear LNA. The performance of the proposed cancellation technique is evaluated through comprehensive RF measurements adopting commercial LTE-Advanced transceiver hardware components. The results show that the proposed technique can provide an additional suppression of up to \textbf{54} dB for the TX passband leakage signal at the LNA input, even at very high transmit power levels and with wide transmission bandwidths. Such novel cancellation solution can therefore substantially improve the TX-RX isolation, hence reducing the requirements on passive isolation and RF component linearity, as well as increasing the efficiency and flexibility of the RF spectrum use in the emerging \textbf{5}G radio networks. 
\end{abstract}


\IEEEpeerreviewmaketitle

\section{Introduction}
\IEEEPARstart{O}{ne} of the fundamental limiting factors in the evolution of wireless communication technologies is the scarcity of radio frequency (RF) spectrum, and consequently finding ways to enhance the spectrum utilization is one of the key elements in existing and emerging radio networks. In current cellular network evolution, spectrum aggregation in the form of contiguous and noncontiguous carrier aggregation (CA) is being adopted to improve the flexibility and efficiency of the radio spectrum utilization \cite{LTE}-\cite{Adnan_Commag}. Furthermore, recently, in-band full-duplex (IBFD) communication has also gained considerable research interest due to its potential to double the spectral efficiency and reduce the communication latency \cite{Sabharwal}, \cite{Bharadia}. 

In all the wireless devices supporting simultaneous transmission and reception, coupling of the own transmit signal into the receiver is one key technical challenge. More specifically, in radio transceivers operating in frequency division duplex (FDD) mode, a duplexer filter is generally used to provide sufficient isolation from the strong transmit signal. However, duplexer filters are generally expensive, bulky, and typically operate in fixed frequency band pairs, making them less attractive for flexible multiband transceivers with low-cost and small form factor. This is further exacerbated with the adoption of CA technology, where due to the reduced duplexing distances achieving sufficient isolation is exceedingly difficult, as acknowledged in 3GPP for both intraband and interband CA \cite{3GPP_duplexer}-\cite{3GPP_duplexer_3}. Meanwhile, suppressing the \emph{self interference} (SI) is the biggest technical challenge in IBFD communications, where the TX signal coupling to the RX can be more than $100$ dB stronger than the desired signal being received concurrently at the same carrier frequency \cite{Bharadia}, \cite{Korpi}.

Using active RF cancellation to complement the passive isolation has been discussed in the literature as one approach to overcome the SI and the TX leakage signal problems in both IBFD and FDD systems \cite{Hong}. In general, the TX leakage signal should be attenuated prior to the RX low noise amplifier (LNA), to prevent the saturation of LNA and analog-to-digital converter (ADC), as well as to avoid RX desensitization. In this context, active RF cancellation solutions with two possible architectures have recently been investigated. The first approach is digitally assisted RF cancellation, where the baseband equivalent of the actual RF cancellation signal is first created in the transceiver's digital front-end through appropriate digital pre-processing of the known transmit data. The corresponding RF cancellation signal is then generated using an auxiliary transmitter chain, and then combined with the received signal in the RX chain \cite{Duarte1}-\cite{Liu2}. Such techniques have the potential of estimating the \emph{coupling channel} response with high accuracy over a wider bandwidth, as majority of the processing is done in the digital domain. However, the cancellation performance is affected by the power amplifier (PA)-induced nonlinear distortion products and the transmitter noise appearing in the RX band \cite{Bharadia}, \cite{Korpi}. Furthermore, existing techniques typically require a separate calibration period where dedicated training signals are transmitted to estimate the coupling channel coefficients. In \cite{Anttila}, a hybrid cancellation scheme is proposed, where the PA nonlinearity-induced intermodulation distortion (IMD) products in the TX leakage signal are mitigated separately in the receiver digital baseband, after an initial RF cancellation phase. Moreover, recently in \cite{Adnan_GlobalSIP} and \cite{Liu2}, the PA nonlinear distortion products are considered in the modeling and cancellation of the transmitter leakage signal already at RF. In \cite{Adnan_GlobalSIP}, Kiayani \emph{et al.} proposed a block least-squares (LS) based approach to effectively estimate the coupling channel in the presence of PA nonlinearities, while Liu \emph{et al.} in \cite{Liu2} proposed a two-step approach where PA nonlinearities are first estimated separately in an observation receiver chain followed by linear coupling channel estimation. In general, these nonlinear cancellation methods provide significant performance improvements. However, none of the existing techniques take into account the potential nonlinear distortion occurring in the LNA during the parameter estimation phase, which can substantially limit the cancellation gain of both linear and nonlinear cancellers due to impaired parameter estimation. For this reason, the existing works typically assume that the LNA is bypassed during the parameter learning, which not only complicates the RX design, but also results in an increased RX noise figure (NF) during the parameter estimation.

The second well-known approach is the pure analog RF cancellation in the transceiver RF front-end, where the PA output signal is used as the input signal to an analog/RF cancellation circuit, that typically consists of delay lines and variable attenuators \cite{Debaillie}-\cite{Zhou}. While these techniques are robust to the transmitter impairments, there are several design concerns and challenges, such as the required number of taps for sufficient TX leakage signal suppression and their optimization, power consumption, and the nonlinear distortion in the cancellation circuitry particularly when the TX power is high \cite{Bharadia}, \cite{Liu}. Furthermore, for MIMO transceivers that are generally equipped with $N_{TX}$ transmit and $N_{RX}$ receive antennas, such an approach will require $N_{TX} \times N_{RX}$ RF canceller circuits to be implemented in contrast to the $N_{RX}$ cancellers needed with the digitally assisted auxiliary transmitter-based approach, thus entailing more cost and complexity to the transceiver design. 
\begin{figure}[!t]
\begin{minipage}[b]{1.0\linewidth}\centering
	\includegraphics[scale = 0.35]{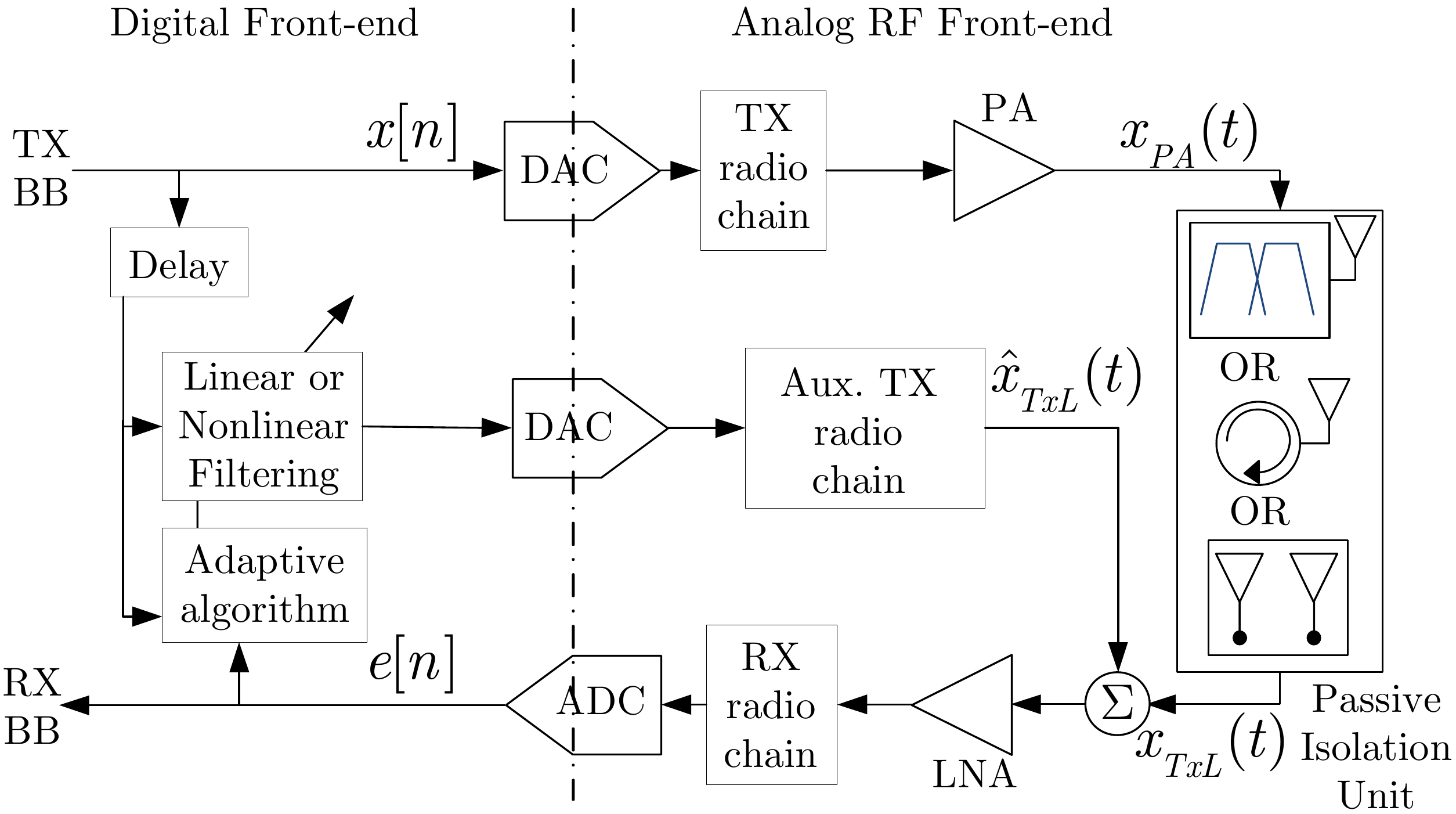}
\end{minipage}
\caption{{Simplified block diagram of a simultaneous transmit-receive radio transceiver employing the active RF cancellation mechanism for suppressing the TX leakage signal. The digital baseband equivalent of the actual RF cancellation signal is first created in the digital domain, while an auxiliary transmitter is utilized to generate the corresponding RF cancellation signal which is finally added to the received signal at the receiver LNA input.}}\label{canc_structure}
\end{figure}

\begin{figure*}[!t]
\begin{minipage}[b]{1.0\linewidth}\centering
	\includegraphics [scale = 1]{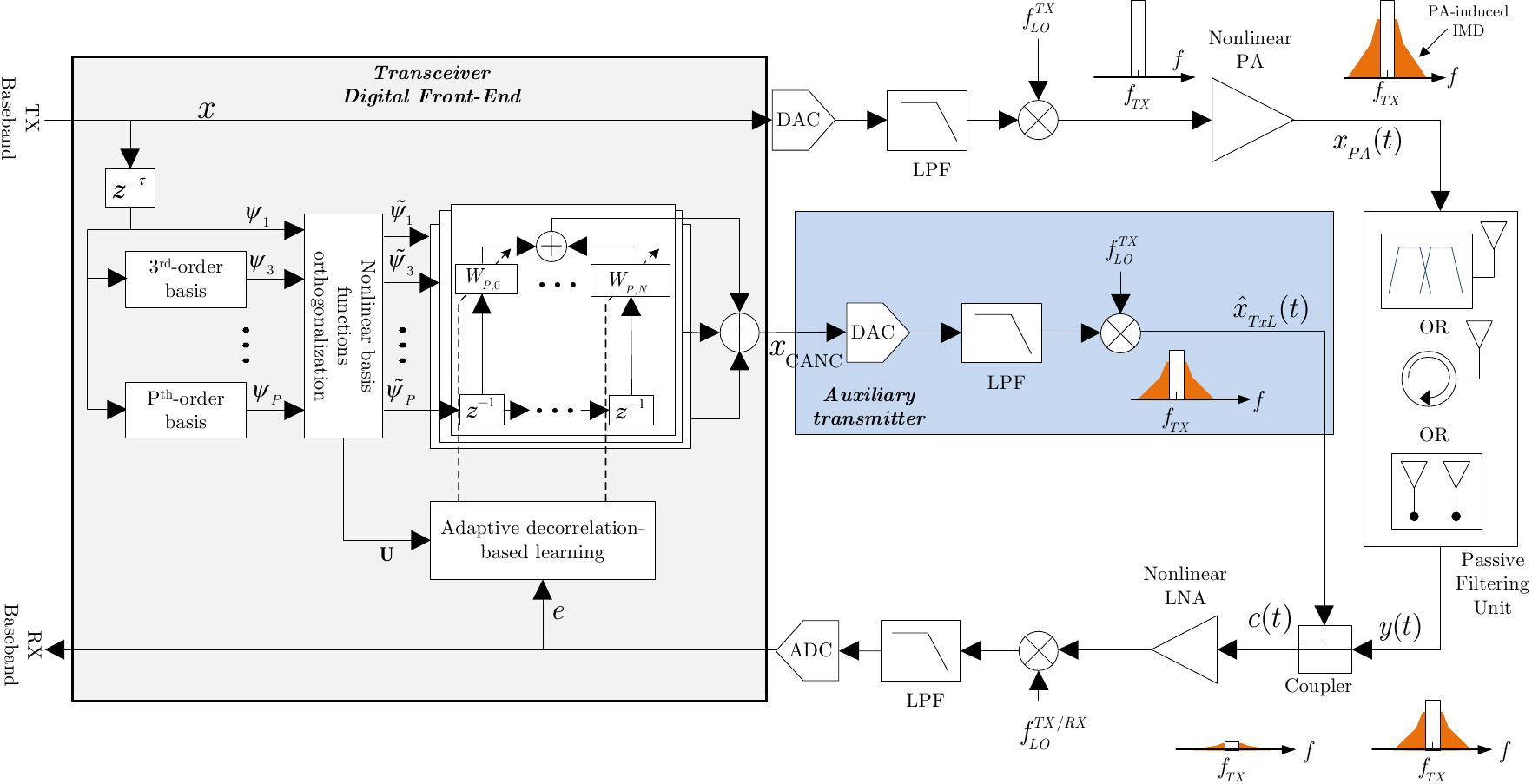}
\end{minipage}
\caption{{A detailed block diagram of a radio transceiver capable of transmitting and receiving simultaneously, incorporating the proposed active RF cancellation approach for the TX passband leakage suppression. The desired received signal, not shown in the figure for simplicity, may be located at the same center frequency as the transmit signal or at a given duplex distance from the transmitter center frequency.}}\label{Main_fig}
\end{figure*}

In this paper, we address the active RF cancellation of the TX passband leakage signals by employing an auxiliary transmitter chain-based approach. It is assumed that some elementary passive isolation is already achieved prior to the active RF cancellation through a duplexer, circulator, or the like, in a shared antenna system, or through proper antenna isolation in a separate antenna system. A block diagram illustrating the considered active RF cancellation concept is shown in Fig.~\ref{canc_structure}. The proposed cancellation technique builds of identifying the nonlinear coupling channel, which models the cascaded response of a nonlinear PA with memory and the potentially frequency-selective passive isolation circuit. Stemming from the nonlinear modeling of the TX leakage signal, developed in our preliminary work in \cite{Adnan_GlobalSIP}, we derive an efficient nonlinear processing structure for the digital processing stage in the overall cancellation path. In addition, we present a novel closed-loop parameter learning approach to estimate the nonlinear cancellation filter coefficients. The proposed decorrelation-based closed-loop learning system targets minimizing the correlation between the nonlinear TX leakage signal at the LNA output and the locally generated nonlinear basis functions. As will be explained in more details below, this approach enables us to avoid bypassing the LNA during the parameter learning, thus keeping the receiver front-end simple while ensuring that the overall RX NF is low. Furthermore, the proposed technique does not require dedicated training signals, and can utilize the online TX data for the parameter estimation. The comprehensive RF measurement results confirm and demonstrate that the proposed nonlinear active RF canceller with closed-loop parameter learning provides substantial suppression of the TX leakage signal, and that the cancellation performance is not essentially degraded as the TX signal power or its bandwidth are increased. Therefore, the proposed technique can significantly improve the overall TX-RX isolation in both FDD and IBFD radio transceivers, and can thus, e.g., simplify the duplexer filter design and the duplex distance requirements in FDD transceivers while also allowing for relaxed RX linearity and dynamic range requirements in IBFD systems.

The remainder of this article is structured as follows. In Section II, we provide the essential signal models for the nonlinear TX leakage signal at the RX input, and building on that, the active RF cancellation solution and the involved nonlinear digital processing stage are formulated. The decorrelation-based closed-loop parameter learning approach is then described in Section III, together with some stability and convergence considerations. Several practical implementation-related aspects for radio transceivers adopting the proposed active RF cancellation solution are discussed in Section IV. The RF measurement results are presented and analyzed in Section V. Finally, the concluding remarks are given in Section VI.

\section{System Model and Active RF Cancellation}
\subsection{Baseband Equivalent Signal Model for Nonlinear TX Passband Leakage}
We begin by developing an overall model of the nonlinear TX passband leakage signal at the LNA input, stemming from a nonlinear TX PA and a finite passive isolation stage, utilizing baseband equivalent signal models and component responses. This model, addressed also in \cite{Adnan_GlobalSIP}, provides the necessary insight into the nonlinear TX leakage signal-induced SI problem in simultaneous transmit-receive systems, and allows us to develop the RF cancellation solution. A block diagram representing a basic radio transceiver architecture together with the corresponding nonlinear TX leakage signal regeneration in the transceiver digital front-end and active RF cancellation structure are shown in Fig.~\ref{Main_fig}. 

Denoting the original baseband transmit signal by $x[n]$, and utilizing the widely-used parallel Hammerstein (PH) model for the TX PA, which is known to provide a good trade-off between accuracy and complexity \cite{Ding}-\cite{Ku}, the baseband equivalent PA output signal can be written as
\begin{equation}\label{PA_op}
	x_{\mathrm{PA}}[n] = \sum_{p=1 \atop p\hspace{1mm} odd}^{P} f_{p,n}\star\underbrace{x[n]\left | x[n] \right |^{p-1}}_{\psi_p[n]}
\end{equation}
where $\psi_p[n] $ denotes the $p^{th}-$order basis function, $P$ is the highest considered PA nonlinearity order, $f_{p,n}$ is the $p^{th}-$order PH branch baseband equivalent filter impulse response modeling the PA memory, and $\star$ represents the convolution operator. In general, the PA nonlinearity generates unwanted IMD products of the transmit signal, resulting in in-band distortion as well as the spectral regrowth around the transmit carrier. 

The PA output signal then propagates towards the antenna through the duplexer or other related passive components, while due to the the finite TX-RX isolation then also partially couples to the RX LNA input. Such TX leakage signal at the receiver input can then be expressed as $x_{\mathrm{TxL}}[n] = h_n \star x_{PA}[n]$ or \cite{Adnan_GlobalSIP}
\begin{equation}\label{TxL}
	x_{\mathrm{TxL}}[n] =\sum_{p=1 \atop p\hspace{1mm} odd}^{P} h_{p,n}\star \psi_p[n],		
\end{equation}
where $h_n$ refers to the basic frequency-selective passive coupling response from the PA output to the LNA input, while $h_{p,n} = h_n \star f_{p,n}$ refers to the corresponding effective coupling channel response for the $p^{th}-$order basis function, which are all assumed to be unknown. 

At the RX LNA input, the desired received signal and the thermal noise are also naturally present, in addition to the TX leakage signal component. Thus, the baseband equivalent total received signal at the LNA input reads 
\begin{equation}\label{RX}
	y[n] = x_\mathrm{D}[n] e^{j\omega_\mathrm{D} n} + x_{\mathrm{TxL}}[n] + \upsilon[n],
\end{equation}
where $x_\mathrm{D}[n]$ represents the desired RX signal, $\upsilon[n]$ refers to the noise, and $\omega_\mathrm{D} = 2\pi \left(f_{\mathrm{TX}} - f_{\mathrm{RX}}\right)/f_s$ represents the normalized duplex distance between the transmitter and receiver carrier frequencies. 

Stemming from the above signal model, we next focus in the following subsection on developing an efficient active RF cancellation solution to suppress the TX leakage.
\subsection{Nonlinear Active RF canceller}
The target of the active RF canceller is to minimize the energy of the TX passband leakage signal component, $x_{\mathrm{TxL}}[n]$ in (\ref{RX}), at the LNA input, such that the SI is suppressed, while also preventing the saturation of the LNA. 

Based on (\ref{TxL}) and (\ref{RX}), perfect cancellation of the nonlinear TX leakage signal can be attained through nonlinear digital filtering or pre-processing of the known transmit data, which effectively incorporates the effects of a nonlinear PA with memory and the frequency-selective passive isolation. Furthermore, a proper delay for synchronous RF cancellation also needs to be applied to the transmit data in the digital baseband. This is followed by the RF up-conversion with an auxiliary transmitter branch, and combining this signal with the received signal at the LNA input. Hence, the baseband equivalent signal at the combiner output, after RF cancellation, can be written as
\begin{equation}\label{cance}
\begin{split}
	c[n] &= y[n] + \hat{x}_{\mathrm{TxL}}[n] \\
			 &= y[n] + h^{\mathrm{Aux}}_n \star \sum_{p=1 \atop p\hspace{1mm} odd}^{P} w_{p,n}\star \psi_p[n-\tau].
\end{split}
\end{equation}
Here, $\tau$ denotes the fixed relative delay between the transmitter leakage signal and the auxiliary transmit path, $w_{p,n}$ is the digital cancellation filter impulse response for the $p$-th order basis function, and $h^{\mathrm{Aux}}_n$ is the overall unknown response of the auxiliary transmitter branch. By substituting (\ref{TxL}) and (\ref{RX}) into (\ref{cance}), the \emph{optimum} pre-processing filters that result in a perfect regeneration of an opposite-phase replica of the nonlinear TX leakage signal and the subsequent RF cancellation, i.e., those that yield $c[n] = x_\mathrm{D}[n] e^{j\omega_\mathrm{D} n} + \upsilon[n]$, can be expressed in frequency-domain as
\begin{equation}\label{comp_filter}
	W^\mathrm{OPT}_p(z) = -\frac{H_p(z)}{H^{\mathrm{Aux}}(z)}; \hspace{9mm} p=1,3,5,\cdots,P.
\end{equation}

In addition to the above cancellation filter parameters, the relative delay $\tau$ needs to be estimated in order to regenerate an accurate replica of the TX leakage. The relative delay is typically static, and can therefore be estimated offline, prior to performing active RF cancellation. The estimation of the cancellation filters, $w_{p,n}$, is, in turn, addressed in the next section, where an adaptive closed-loop learning system based on the decorrelation principle is proposed.

\section{Closed-Loop Parameter Learning}
The derived optimum cancellation filters in (\ref{comp_filter}) depend on the linear and nonlinear coupling channel responses of different orders and the auxiliary transmitter response, all of which are unknown. Hence, for high-accuracy cancellation, these responses must be estimated, explicitly or implicitly. For computing friendly but efficient estimation processing, we develop in this section a closed-loop solution where the digital cancellation filters, $w_{p,n}$ in (\ref{cance}), are iteratively adapted, while observing the LNA output at TX passband, to minimize the TX leakage power at LNA input. Compared to the existing parameter estimation solutions, e.g., in \cite{Duarte1}-\cite{Liu2}, which are known to suffer from nonlinear distortion due to LNA in the learning phase, the proposed closed-loop approach is substantially more robust in this respect while also offering a reduced computing complexity. This will be clearly demonstrated and verified through the extensive RF measurement results in Section V. Furthermore, the iterative or adaptive estimation processing also facilitates tracking any possible variations in the coupling channel characteristics due to the changes, e.g., in the PA characteristics, or passive isolation circuits. 

We begin by shortly introducing first the basis function orthogonalization procedure, and then describe the actual iterative closed-loop parameter learning algorithm adopting the so-called decorrelation principle. In general, it is pertinent to note that in IBFD systems where transmitter and receiver are operating at the same carrier frequency, the nonlinear TX passband leakage signal is inherently at the main RX passband and can thus be observed through it. However, in the FDD case with TX and RX tuned to different center frequencies, a separate observation receiver chain may be needed for observing the nonlinear TX leakage signal in the parameter learning context. We elaborate further on this issue and discuss various RX chain implementation alternatives in subsection IV-C of the paper.

\subsection{Basis Function Orthogonalization}
In general, the nonlinear basis functions of different orders, $\psi_p[n] = x[n]\left|x[n]\right|^{p-1}$, are strongly \emph{mutually correlated}. As a result, adaptive learning algorithms will suffer from slow convergence and potentially high excess mean square error, thus limiting the cancellation performance. To ensure faster and smoother learning and high cancellation performance, as well as better numerical properties in digital hardware implementations, the basis functions can be first \emph{orthogonalized} with respect to each other - an approach that is widely adopted in SI cancellation and digital pre-distortion (DPD) processing contexts in general \cite{Korpi2}, \cite{Mahmoud_TMTT}. To shortly outline the orthogonalization procedure, we switch to vector-matrix notations, and collect the instantaneous delayed basis function samples in a vector as
\begin{equation}\label{Psi_vect}
	\Psi[n-\tau]=\begin{bmatrix}
\psi_1[n-\tau] &\psi_3[n-\tau]  &\cdots  & \psi_P[n-\tau]
\end{bmatrix}^T.
\end{equation}
Then, a new vector of instantaneously orthogonalized basis function samples, denoted by $\tilde{\Psi}[n]$, is generated as
\begin{equation}\label{Ortho_Basis}
	\tilde{\Psi}[n-\tau] = \textbf{S} \Psi[n-\tau],
\end{equation}
where $\textbf{S}$ denotes the transformation matrix. The transformation matrix can be calculated through, e.g., singular value or QR decomposition \cite{Mahmoud_TMTT}, \cite{Haykin}, or alternatively using the eigen decomposition of the covariance matrix of nonlinear basis functions, as described in \cite{Korpi2}. The latter approach is in general beneficial from implementation point of view, as the corresponding transformation matrix depends only on the statistical properties of the transmit signal and thus does not need to be evaluated for each individual TX data symbol or block. However, the transformation matrix must be recomputed and updated when the statistics of the transmit signal changes at large, i.e., when completely changing the radio access technology. Thus, for a given radio access technology, it can be pre-computed offline, and stored locally.

\subsection{Block-Adaptive Learning Algorithm Through Decorrelation}
To minimize the TX leakage at LNA input, the proposed closed-loop learning algorithm is based on minimizing the correlation between the current baseband observation of the nonlinear TX passband leakage signal and the basis function samples, constructed from the known baseband transmit data with the proper delay. Such decorrelation-based learning concept was adopted by the authors in \cite{Mahmoud_TMTT}, in the context of DPD, while is now being deployed as an efficient means for RF canceller parameter estimation. While the context and application are here different, we acknowledge that computing-wise there is clear similarity to our earlier work in \cite{Mahmoud_TMTT}.

Now, we assume that the digital cancellation filter length is $N+1$ per nonlinearity order, and the estimation block size is $M$ samples per learning iteration. Then, the cancellation filter coefficients are updated using the block-adaptive decorrelation-based algorithm as
\begin{equation}\label{Decorr_coeff}
	\textbf{w}[m+1] = \textbf{w}[m] - \mu \left[\textbf{e}[m]^H \textbf{U}[m] \right]^T,
\end{equation}
where $\mu$ denotes the learning step-size and the superscript $(.)^H$ denotes the Hermitian transpose. In the above equation, the utilized samples and the corresponding cancellation filter coefficients, within the processing block $m$, are collected into the following vectors and matrices as
\begin{equation}\label{mtx_vect}
\begin{split}
\textbf{w}_p[m] &=
\begin{bmatrix}
w_{p,0}[m] & w_{p,1}[m] & \cdots & w_{p,N}[m]
\end{bmatrix}^T \\
\textbf{w}[m] &=
\begin{bmatrix}
\textbf{w}_1[m]^T & \textbf{w}_3[m]^T & \cdots & \textbf{w}_P[m]^T
\end{bmatrix}^T \\
{\textbf{u}}_p[n_m] &=
\begin{bmatrix}
{\tilde{\psi}}_p[n_m] & {\tilde{\psi}}_p[n_m-1] & \cdots & {\tilde{\psi}}_p[n_m-N] 
\end{bmatrix}^T \\
\textbf{U}_p[m] &=
\begin{bmatrix}
{\textbf{u}}_p[n_m] & {\textbf{u}}_p[n_m+1] & \cdots & {\textbf{u}}_p[n_m+M-1] 
\end{bmatrix}^T \\
\textbf{U}[m] &=
\begin{bmatrix}
\textbf{U}_1[m] & \textbf{U}_3[m] & \cdots & \textbf{U}_P[m]
\end{bmatrix} \\
\textbf{e}[m] &= 
\begin{bmatrix}
e[n_m] &e[n_m+1] &\cdots  &e[n_m+M-1] 
\end{bmatrix}^T.
\end{split}
\end{equation}
Here, $\textbf{w}_p[m]$ denotes the current cancellation filter impulse response, of dimension $(N+1) \times 1$, corresponding to $p^{th}-$orthogonalized basis function, while the aggregate filter $\textbf{w}[m]$ of dimension $((P+1)/2)(N+1) \times 1$ stacks all parallel filters together. In addition, at learning iteration $m$, $\textbf{U}[m]$ is an aggregate data matrix of size $M \times ((P+1)/2)(N+1)$ collecting all the transformed orthogonalized basis function samples of different orders into a single matrix, and is composed of sub-matrices $\textbf{U}_p[m]$ that are all of size $M \times (N+1)$, and $n_m$ denotes the index of the first sample of the processing block $m$. Notice from Fig.~\ref{Main_fig} that the physical cancellation is performed with a coupler in the RF domain at the LNA input, whereas, the error signal used in the parameter learning is the true baseband observation of the cancelled signal after having propagated through the LNA, down-conversion and filtering stages. Thus, in equation (\ref{Decorr_coeff}), $\textbf{e}[m]$ denotes the vector of observed error signal samples of size $M \times 1$ that contains the true baseband samples of the nonlinear TX passband leakage signal observed under the current canceller filter parameters $\bold{w}[m]$. Finally, the aggregate output vector of the digital cancellation filters, of size $M \times 1$, for the processing block $m$ is given by
\begin{equation}\label{blk_LMS_op}
	\textbf{x}_{\mathrm{CANC}}[n] = \textbf{U}[m]\textbf{w}[m].
\end{equation}

In general, we acknowledge that from the computational perspective, the block decorrelation-based learning algorithm in (\ref{Decorr_coeff}) is essentially similar to the widely-known block least mean squares (LMS) adaptive filtering principle \cite{Clark}, \cite{Haykin}. The different sign in the update rule in (\ref{Decorr_coeff}), compared to block LMS, is stemming from the fact that the RF cancellation is assumed to reflect addition, instead of subtraction, which can easily be shown to change the sign of the gradient of the absolute squared error.

In general, it is useful to note that the amount of learning samples $M$ utilized within block $m$ can be chosen independently of the actual transmit data sequence length $L$, and commonly $M << L$. In addition, the closed-loop learning system does not call for any specific training or pilot signals but the actual online transmit data can be directly utilized. Moreover, the speed of convergence and the residual power of the TX leakage signal are affected by the choice of the step-size value. Since different orders of basis functions may have different powers, a different step-size value can be applied for each basis function. However, in this paper, we use the same step-size value for all basis functions, while further investigating and finding a set of step-size values for each basis function is an important future work item. 

Finally, notice that in the very beginning of the parameter learning, the TX leakage power can be very high and thus the LNA output observation contains additional LNA-induced nonlinear distortion, which degrades the parameter estimation accuracy of existing reference solutions \cite{Duarte1}-\cite{Liu2}. However, since the proposed parameter learning is a closed-loop system, the update algorithm will steer the coefficients towards a solution where the leakage power at the LNA input starts to reduce. This, in turn, reduces the LNA-induced nonlinear distortion, and the proposed system will converge towards a state where the leakage power, and therefore also the LNA-induced distortion, are essentially minimized. In the next sub-section, we address this more rigorously in terms of loop stability.

\subsection{Stability Analysis}
In order to characterize the stability of the proposed RF canceller and closed-loop learning system in a more rigorous manner, let us define the limits for the step size parameter $\mu$. Noting that the proposed closed-loop learning system is indeed essentially identical to the block least mean squares (LMS) approach when it comes to the specific computing algorithm in the parameter learning stage, the well-known results regarding the block LMS algorithm can also be applied here. For this, the aggregate filter input data vector at an arbitrary time instant $n$ is first defined as
\begin{equation}\label{LMS_in}
\textbf{u}[n] = \begin{bmatrix} {\textbf{u}}_1[n]^T & {\textbf{u}}_3[n]^T & \cdots & {\textbf{u}}_P[n]^T \end{bmatrix}^T
\end{equation}
This is in fact one transposed row of the total input matrix~$\textbf{U}[m]$ in (\ref{mtx_vect}). Assuming then first that the LNA is not significantly distorting the cancelled signal, the convergence of the coefficients is only dependent on the statistics of the aggregate filter input data in (\ref{LMS_in}), as is well known in the existing LMS literature. In particular, as shown in \cite[p. 450]{Haykin}, the stability of the proposed closed-loop learning system is ensured when
\begin{equation}\label{mu_range}
0 < \mu < \frac{2}{M \lambda_\mathrm{max}(\mathbf{R})}
\end{equation}
where $\lambda_\mathrm{max}(\textbf{R})$ is the largest eigenvalue of the correlation matrix of the aggregate filter input vector, defined as $\textbf{R} = \operatorname{E}\left[\textbf{u}[n] \textbf{u}[n]^H\right]$. In other words, the stability of the canceller is ensured by choosing a suitably small value for the step size $\mu$. This limit is also taken into consideration in the RF canceller and closed-loop learning system implementation reported in Section V, as evidenced by the high cancellation performance.

When it comes to the learning system stability under the nonlinear operation region of the LNA, scenarios where the error signal is nonlinearly distorted have also been investigated in the earlier LMS-based adaptive filtering literature. Perhaps the most extreme case is the hard limiter, where only the sign of the error signal is used for parameter learning \cite[p. 135]{Diniz97}. This type of an LMS-variant is usually referred to as the \emph{sign-error algorithm}, and it essentially corresponds to a fully saturated nonlinearity in our physical closed-loop system context. Therefore, it can be considered a very pessimistic model for the nonlinear LNA, reflecting a case where the input power of the LNA is overly high. In \cite[p. 138]{Diniz97}, a stability condition for this type of an LMS algorithm is derived in closed form, which clearly proves that a suitably small step size does indeed ensure the stability, even when the error signal is distorted in a heavily nonlinear manner. For brevity, we do not numerically evaluate the step size limit here, and instead we kindly ask the reader to refer to \cite[pp. 135-144]{Diniz97} for further information.

In addition, the effect of a generic saturating nonlinearity in the feedback loop of LMS-type algorithms is also investigated in \cite{Bershad88}. There, a Gauss error function is used to model the nonlinearity that is distorting the error signal. This type of a function is a saturating nonlinearity and, with proper parametrization, it closely resembles the behavior of a nonlinear amplifier, such as an LNA. The convergence behavior of the LMS under such a smooth nonlinearity is analytically investigated in \cite{Bershad88} and the obtained results again show that, by choosing a suitably small step size, the coefficients do indeed converge in a stable manner. Hence, the findings in \cite{Diniz97} and \cite{Bershad88} provide a solid basis for more rigorously concluding that the proposed RF canceller and closed-loop learning system are stable even under a nonlinear LNA, as long as the step size is chosen accordingly. For brevity, the exact boundaries for the step size are not discussed herein, while the step sizes used in the results reported in Section V are obviously chosen such that the stability of the closed-loop system is guaranteed.

\section{Transceiver Implementation Aspects}
In this section, we shortly review the implications of adopting the proposed active RF cancellation solution on the transceiver design and its operation. 

\subsection{Analysis of TX Noise at RX Band}
As highlighted earlier, the adopted auxiliary transmitter based active RF cancellation structure cannot account for the TX noise, while the auxiliary TX may also contribute to the total effective noise floor seen by the receiver. In the following, we analyze and characterize the TX noise aspects and its impact on the receiver performance through transceiver system calculations. 

In general, on the TX side, in addition to the thermal noise floor, there is also quantization noise present in the transmit signal, produced by the digital-to-analog converter (DAC). Therefore, the total transmitter noise power, on a linear scale, can be defined as
\begin{equation}\label{TX_noise}
	p_{\mathrm{noise}}^{\mathrm{TX}} = g^{\mathrm{TX}} \left(F^\mathrm{TX}p_{\mathrm{noise}}^{\mathrm{Thermal}} + p_{\mathrm{noise}}^{\mathrm{Quant}}\right),
\end{equation}
where $g^{\mathrm{TX}}$ is the total gain of the TX chain, $F^\mathrm{TX}$ is the noise factor of the TX, while $p_{\mathrm{noise}}^{\mathrm{Thermal}}$ and $p_{\mathrm{noise}}^{\mathrm{Quant}}$ are the thermal noise power and quantization noise power of the DAC, respectively. The quantization noise power density per Hz of the DAC can be expressed as
\begin{equation}\label{Quan_Noise}
\begin{split}
	P_{\mathrm{noise}}^{\mathrm{Quant}} &= P^\mathrm{Avg}_\mathrm{DAC} - SNR_\mathrm{DAC} \\
												 &= P^\mathrm{Avg}_\mathrm{DAC} - 6.02b - 4.76 + \mathrm{PAPR} - 10\log_{10}\left(f_s/2\right),
\end{split}
\end{equation}
where $P^\mathrm{Avg}_\mathrm{DAC}$ denotes the average power of the signal at DAC output, $b$ denotes the number of bits in the DAC, $\mathrm{PAPR}$ is the peak-to-average power ratio of the transmit waveform, and $f_s$ represents the sampling frequency. The last term in the above equation represents the processing gain, i.e., improvement in signal-to-noise ratio (SNR) of the signal due to oversampling.

At the receiver LNA input, the overall TX-induced noise is composed of the main transmitter's noise being further suppressed by the passive isolation, $\alpha_\mathrm{iso}$, and the noise of the auxiliary transmitter chain coupling into the main receive path. Thus, the total TX-induced noise power at the receiver input can equivalently be expressed as
\begin{equation}\label{TxL_noise}
	p_{\mathrm{noise}}^{\mathrm{TX-induced}} = \alpha_\mathrm{iso}p_{\mathrm{noise}}^{\mathrm{TX,Main}} + Cp_{\mathrm{noise}}^{\mathrm{TX,Aux}},
\end{equation}
where $C$ denotes the coupling factor of the directional coupler at LNA input. 

Now, to provide numerical results, we consider IBFD transceiver operation where transmitter noise is more challenging to handle compared to classical FDD operation, and assume typical transceiver component values corresponding to cellular mobile devices. The PAPR of the transmit signal is assumed to be $7$ dB, the maximum average power of the DAC is $-6$ dBm, the number of DAC bits is $14$, the sampling frequency is $30.72$ MHz, and the fundamental thermal noise density is $-174$ dBm/Hz. The gain of the main transmitter chain is assumed to be $29$ dB, and its noise figure is assumed to be $10$ dB. For the auxiliary transmitter chain, the gain and noise figure are assumed to be $5$ dB and $9$ dB, respectively. The signal from the main transmit path is attenuated by a certain amount of passive isolation, and a realistic passive isolation of $40$ dB is assumed here as a baseline number. The coupling factor of the directional coupler is assumed to be $-15$ dB, which ensures in this particular example that the TX leakage signal from the main transmitter path and the cancellation signal from the auxiliary transmitter chain will have approximately equal powers at the LNA input. Then, using the above formulas, the power of the main transmitter noise leaking into the receiver is at $-169.5$ dBm/Hz, while the auxiliary transmitter noise power at the receiver input is $-168.7$ dBm/Hz. The total power of the TX-induced noise at the receiver input can thus be computed to be $-166$ dBm/Hz. When compared to the receiver thermal noise power, which is at $-170$ dBm/Hz when assuming $4$ dB RX NF, this analysis indicates that the transmitter noise in auxiliary transmitter chain-based architectures may reduce the receiver sensitivity in simultaneous transmit-receive radios. However, it should be noted that its impact can be reduced through proper RF design and that the impact is not extensively large. 
\begin{figure}[!t]
\begin{minipage}[b]{1.0\linewidth}\centering
	\includegraphics[scale = 0.51]{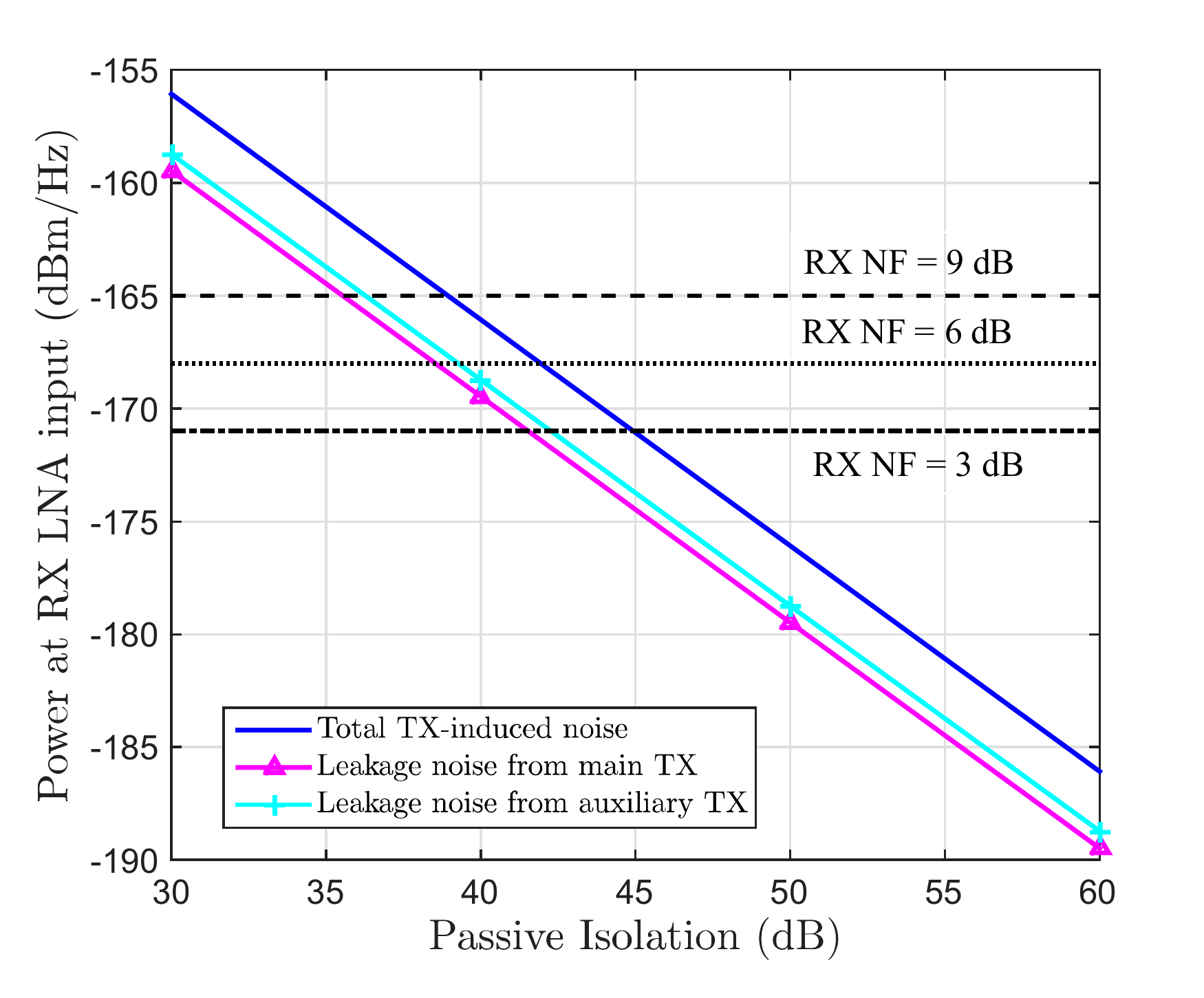}
\end{minipage}
\caption{The power levels of different TX noise components at the input of RX LNA as a function of passive isolation. The reference receiver thermal noise powers with different RX NFs are also shown.}\label{TX_noise}
\end{figure}

Next, in order to obtain further insight on the noise aspects, the power levels of the different TX noise components at the input of RX LNA are evaluated against different passive isolation levels and plotted in Fig.~\ref{TX_noise}, together with the reference receiver thermal noise powers assuming different RX NFs. Here, the gain of the auxiliary RX chain is appropriately adjusted such that the cancellation signal from the auxiliary transmitter chain and the leakage signal have similar powers at the RX LNA input. It is obvious from the figure that with low passive isolation, the transmitter noise can indeed impair the receiver sensitivity, whereas if one assumes the UE receiver noise figure to be higher, say 9 dB as assumed in 3GPP standardization \cite{LTE_UE}, then the excess noise impact is vanishingly small with typical passive isolation levels.

\subsection{RX Noise Figure Aspects}
As shown in Fig.~\ref{Main_fig}, a directional coupler is employed, instead of a power combiner, to add the cancellation signal to the received signal. From the weak received signal perspective, the adoption of a combiner, particularly at the LNA input, is generally not seen feasible as a combiner would introduce additional insertion loss, and would also contribute to the overall receiver noise figure. However, compared to power combiners, the coupler only slightly degrades the receiver noise figure. For instance, compared to a $3$ dB degradation in NF due to a power combiner, a $10$ dB directional coupler increases the RX noise figure only by $0.4$ dB. Therefore, the proposed TX leakage signal cancellation architecture with a directional coupler imposes only a small penalty in the overall RX noise figure. 

\subsection{RX Chain Implementation Considerations}
In the parameter estimation phase, the TX passband leakage signal at LNA interface must be observed. This observation can be extracted either by using the device's main receiver or through a separate observation receiver. In the IBFD case, both the transmitter and receiver local oscillators are tuned to the same center frequency, therefore the main receiver can be directly used for observing the TX leakage signal. However, in the specific case of FDD transceivers where the transmitter and the receiver are operating on different carrier frequencies, the main receiver cannot automatically be used for extracting the transmitter passband leakage signal, and subsequently for the cancellation filter parameter estimation. Therefore, a dedicated observation receiver chain, already commonly present in radio transceivers for DPD parameter estimation purposes, can potentially be utilized by switching its input from the PA output to the LNA output. Alternatively, it can be argued that the main RX can also be tuned momentarily to the TX frequency in order to sense the TX leakage signal. The latter approach is also realistic as the PA nonlinearity and the coupling channel characteristics are generally slowly varying, therefore the parameter estimation can be performed offline or regularly at dedicated calibration phases. It is acknowledged that if the main RX is  momentarily configured to observe the TX passband frequencies in FDD systems, no useful signals can be received during such reconfiguration period.

\subsection{Computational Complexity Analysis}
In this subsection, we present the computational complexity analysis of the proposed technique, evaluated in terms of the floating point operations (FLOP) \cite{Tehrani}. In general, the complexity of the proposed technique consists of two parts, namely the processing complexity to regenerate the baseband equivalent replica of the self-interference under given filter coefficients, and the complexity of the cancellation filter parameter learning. The self-interference regeneration complexity is further composed of three parts - the complexity of the basis function generation, the basis function orthogonalization, and the basis function filtering. By adopting the notations used throughout this paper and assuming that, in total, $B$ blocks are utilized for the parameter learning, i.e., $m = 1, 2, \cdots, B$, the results of the complexity analysis for the proposed technique are summarized in Table~\ref{comp_complx}, where it has been assumed that the transformation matrix values are pre-computed.

In the next section, we evaluate and present concrete numerical values of the running complexity of the proposed self-interference regeneration, in terms of giga FLOP per second (GFLOP/s), and the parameter learning complexity, in terms of mega FLOP (MFLOP) per the overall learning procedure, corresponding to the RF cancellation performance results.

\begin{table*}[t]
\small
\centering
\caption{The computational complexities of the proposed self-interference regeneration and parameter learning stages.}
\label{comp_complx}
\begin{tabular}{c|c|c|c|c}
\toprule
\multicolumn{4}{c|}{\begin{tabular}[c]{@{}c@{}} Self-interference regeneration complexity \\(FLOP/sample)\end{tabular}}                                                                                                                                                                                                                                              & \multirow{3}{*}{\begin{tabular}[c]{@{}c@{}}Cancellation filter parameter \\learning complexity\\ (FLOP/$BM$ samples)\end{tabular}} \\ \cline{1-4}
\multirow{2}{*}{\begin{tabular}[c]{@{}c@{}}Basis function \\generation\end{tabular}} & \multicolumn{2}{c|}{\begin{tabular}[c]{@{}c@{}}Basis function orthogonalization\end{tabular}}                                                                                       & \multirow{2}{*}{\begin{tabular}[c]{@{}c@{}}Basis function\\ filtering\end{tabular}} &                                                                                                                                   \\ \cline{2-3}
                                                                                      & \begin{tabular}[c]{@{}c@{}} QR \\decomposition-based\end{tabular} & \begin{tabular}[c]{@{}c@{}}Covariance matrix eigenvalue \\decomposition-based\end{tabular} &                                                                                      &                                                                                                                                   \\ \hline
$P+2$                                                                                 & $(P+1)^2 + 2(P+1)$                                                                             & $2(P+1)^2$                                                                             & $4(P+1)(N+1)-2$                                                                      & $B(P+1)(N+1)(4M+1)$                                                                                                               \\ 
\bottomrule
\end{tabular}
\end{table*}

\begin{figure*}[!t]
\begin{minipage}[b]{1.0\linewidth}\centering
	\subfigure[]{\label{PSD_10MHz}\includegraphics [scale = 0.24]{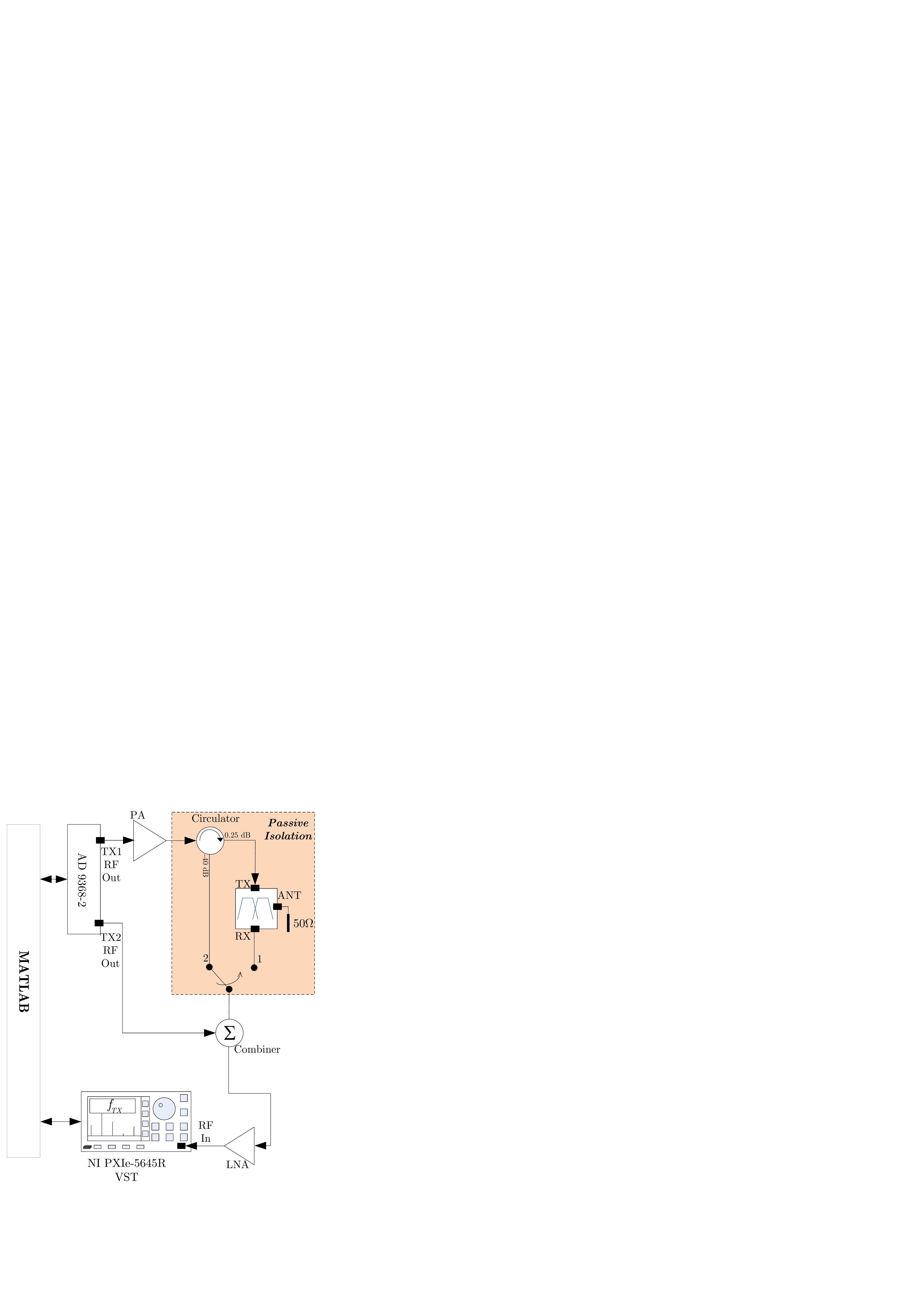}}
	\subfigure[]{\label{PSD_20MHz}\includegraphics [scale = 0.075]{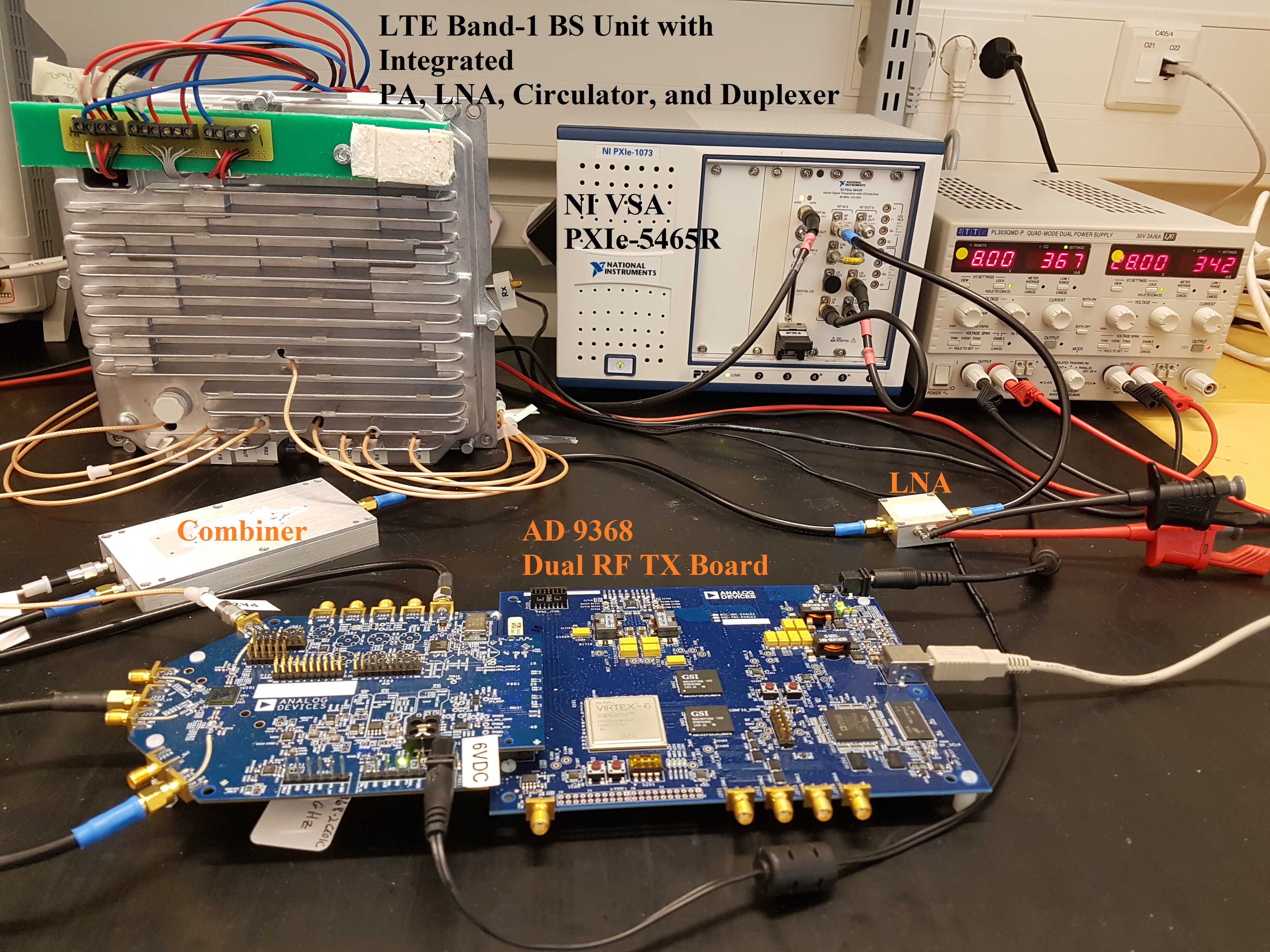}}
\end{minipage}
\caption{{The RF measurement setup used for evaluating the performance of the proposed nonlinear active RF cancellation solution: (a) block diagram of the measurement setup; (b) photo of the measurement setup. }}\label{Meas_setup}
\end{figure*}

\section{RF Measurement Results}
\subsection{Measurement Setup and Parameters}
In this section, we report RF measurement results to demonstrate and verify the high cancellation performance of the proposed technique. The measurement setup is shown in Fig.~\ref{Meas_setup}, and the measurements are carried out by adopting LTE-Advanced Band $1$ base station hardware (downlink: 2110-2170 MHz), namely, the PA, duplexer, circulator, and LNA modules. In addition, in the setup, the Analog Devices evaluation board (model no. AD9368-2), which is equipped with two RF transmitter chains, is used to implement the main and auxiliary transmit paths. The output of the first transmitter chain from the evaluation board is fed to a commercial BS PA (model no. MD7IC2250GN), which has $31$ dB gain and $+47$ dBm $1-$dB compression point. The PA output is connected to a circulator, and is followed by a duplexer. The circulator has $0.25$ dB insertion loss in the forward direction and $40$ dB isolation, whereas the duplexer has frequency-selective $70-72$ dB isolation for the TX passband frequencies. Therefore, a strong nonlinear TX signal is indeed leaking into the RX chain. In the RX chain, a combiner is used to inject the cancellation signal coming from the second transmitter chain of the evaluation board to the LNA input. The signal at the combiner output is amplified by a LNA, which is then fed to the RF input of the National Instrument (NI) PXIe-5645R vector signal transceiver (VST). The VST has an effective capture bandwidth of $61$ MHz and receiver sampling rate of $120$ MHz rate, and it is used here for down-conversion and digitization of the received signal. A host processor equipped with MATLAB is used for performing the DSP related tasks, as well as to control all the measurement instruments. 

\begin{figure*}[!t]
\begin{minipage}[b]{1.0\linewidth}
\centering
	\subfigure[]{\label{PSD_10MHz}\includegraphics [scale = 0.5]{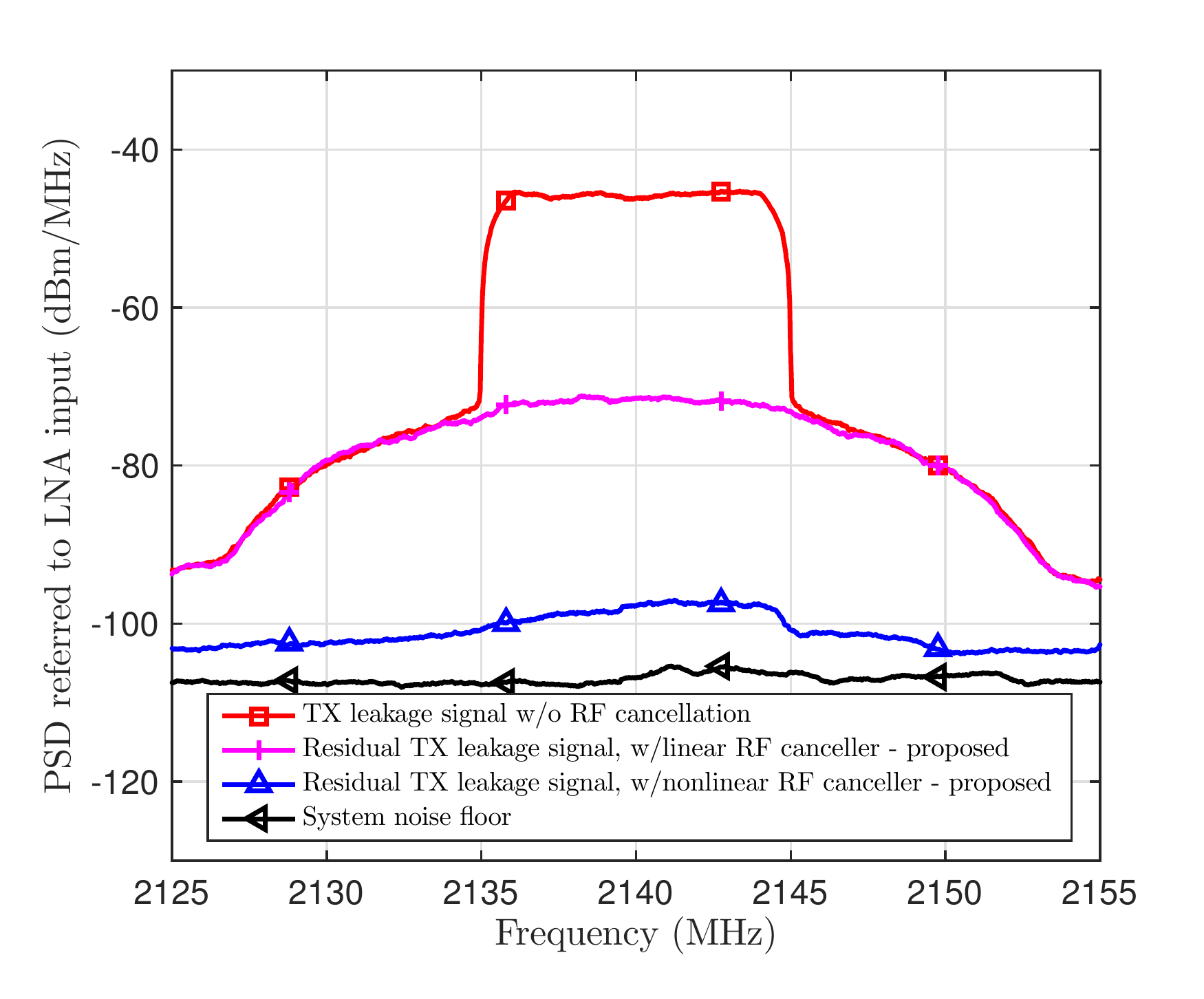}}
	\subfigure[]{\label{PSD_20MHz}\includegraphics [scale = 0.5]{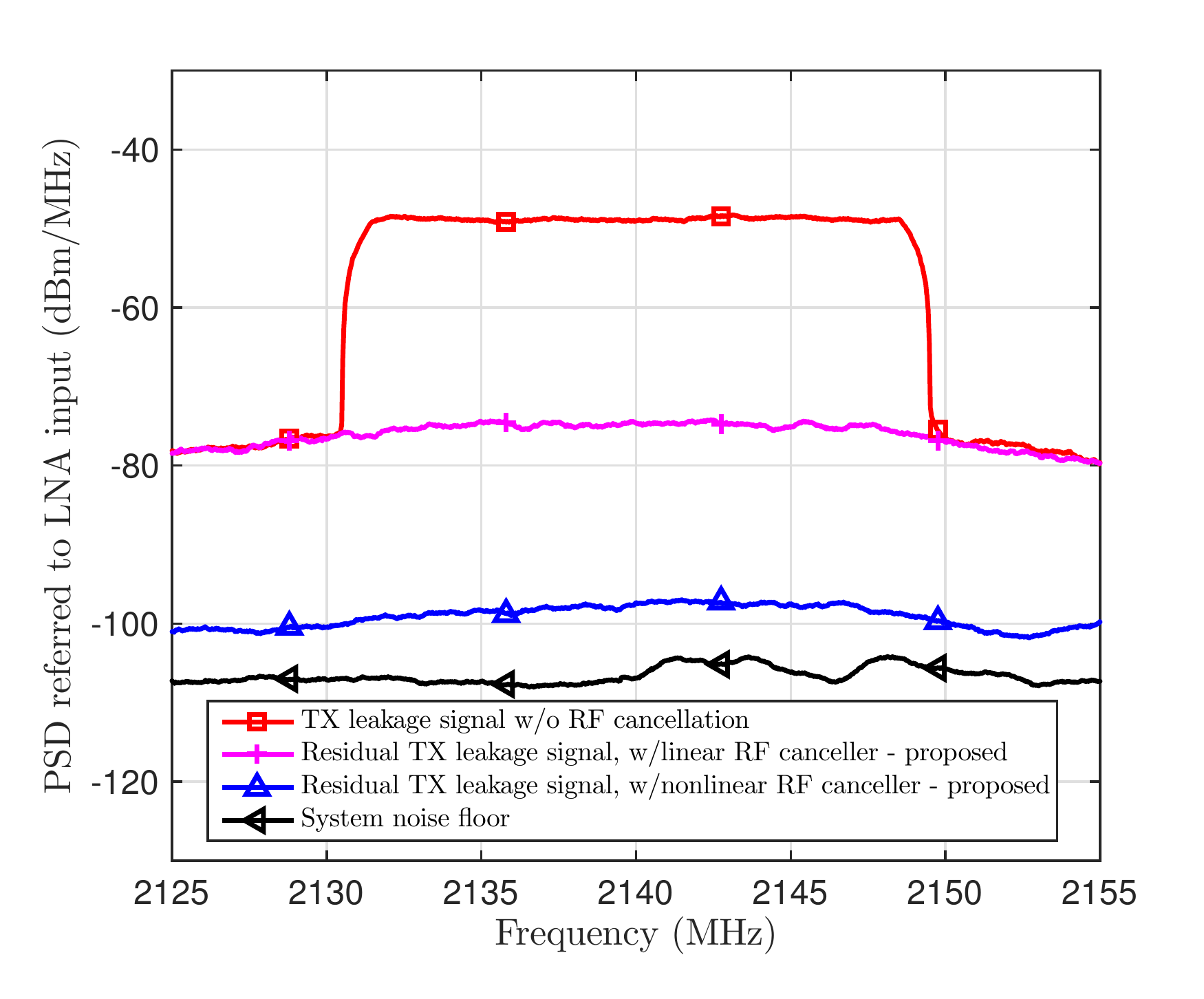}}
	\end{minipage}
\caption{Measured signal spectra of the nonlinear TX leakage signal before and after different RF cancellers for a duplexer-based FDD transceiver experiment. TX power is $+35$ dBm, the average duplexer isolation at TX passband frequencies is $72$ dB, LNA out-of-band IIP3 is 0 dBm, and the TX center-frequency is $2.14$ GHz: (a) TX signal bandwidth: 10 MHz; (b) TX signal bandwidth: 20 MHz.}\label{Meas_PSD}
\end{figure*}

In all the measurements, the transmit signal is an LTE-Advanced downlink signal with $16$-QAM subcarrier modulation and $8$ dB PAPR, where iterative clipping and filtering based PAPR reduction approach \cite{Armstrong} is applied to the transmit signal. The TX center frequency is $2140$ MHz. Furthermore, a proper relative delay is applied to the auxiliary transmit path signal such that the cancellation signal and the transmitter leakage signal are time aligned at the combiner input for synchronous cancellation. Since the relative delay is static, it is therefore estimated only once prior to the RF cancellation by transmitting \emph{frequency-interleaved} orthogonal signals simultaneously on both main branch and auxiliary branch transmitters. Then, by utilizing the composite baseband received signal and the original transmit data, the relative delay is computed in the digital domain, and is stored locally. The block size for the cancellation filter parameter estimation is $M = 13. 000$, the parameter estimation sampling rate is $61.44$ MHz, and the total number of block-adaptive iterations is $B = 25$. Moreover, the cancellation filter coefficients are estimated with a randomly-drawn transmit signal realization, and after the proposed closed-loop learning system has converged, the actual achievable RF cancellation performance is evaluated using another randomly-drawn transmit signal realization. Both FDD and IBFD scenarios are measured and reported, while we explicitly assume that no received signal-of-interest is present. This is purposely done in order to assess the absolute performance of the proposed RF canceller. Furthermore, for comparison purposes, both linear canceller ($P = 1$) and nonlinear canceller ($P \geq 3$) are experimented and measured.
\subsection{Duplexer-Based Measurement Results}
The cancellation performance is first experimented and evaluated with a duplexer setup, corresponding to a FDD transceiver operation, where two different bandwidth scenarios are considered, i.e., $10$ MHz and $20$ MHz transmit signals. The transmit power is $+35$ dBm and after duplexer isolation the TX leakage signal power at the LNA input, without RF cancellation, is $-37$ dBm. Note that in this considered FDD transceiver case, the main receiver is assumed to observe the LNA output at TX center-frequency during parameter learning, thus it is not possible to receive any actual useful received signal. The adopted LNA (model no. MGA-14516) has $31$ dB gain and out-of-band IIP3 of 0 dBm. The adopted nonlinear canceller order is $P = 7$, and the compensation filter length per nonlinearity order is $9-$taps for $10$ MHz transmit signal bandwidth, and $13-$taps for $20$ MHz transmit signal bandwidth. Fig.~\ref{Meas_PSD} shows the LNA input-referred spectra of the measured nonlinear transmitter leakage signal before and after active RF cancellation, whereas the achieved self-interference suppression and the corresponding computational complexity are summarized in Table~\ref{comp_FDD}. We observe that without active RF cancellation, the TX leakage is some $60$ dB above the receiver noise floor, reflecting parameter estimation thermal noise SNR. In general, we can deduce two important results from the PSD curves in Fig.~\ref{Meas_PSD}: first, the cancellation performance of the proposed linear canceller is limited due to the presence of strong PA-induced nonlinear distortion products in the TX leakage signal while the proposed nonlinear RF canceller is indeed capable of efficiently suppressing the nonlinear TX leakage signal close to the system noise floor, in particular when a proper set of nonlinearity orders is used in the modeling and digital baseband regeneration of the RF cancellation signal. Second, the cancellation performance is not heavily degraded as the transmit signal bandwidth increases, because the cancellation filter parameter estimation and the leakage signal regeneration are done in the digital domain. Complexity-wise, as can be observed in Table II, the actual interference regeneration clearly dominates over the parameter learning. Overall, the involved processing complexity in the range of few tens of GFLOP/s is clearly within the processing capabilities of modern base-stations.

\begin{table}[]
\small
\centering
\caption{Comparison of the nonlinear TX leakage signal suppression with proposed active RF cancellation technique in a duplexer-based FDD transceiver experiment. TX power is $+35$ $\mathrm{dBm}$, average duplexer isolation is $72$ $\mathrm{dB}$, and the transmit signal bandwith is $20$ $\mathrm{MHz}$. Also the involved processing complexities are shown.}
\label{comp_FDD}
\begin{tabular}{l|c|c|c}
\toprule
\multirow{2}{*}{}                                                                           & \multirow{2}{*}{\begin{tabular}[c]{@{}c@{}}\\ Power\\ (dBm)\end{tabular}} & \multicolumn{2}{c}{Complexity}                                                                                                                          \\ \cline{3-4} 
                                                                                            &                                                                        & \begin{tabular}[c]{@{}c@{}}Interference\\ regeneration\\ (GFLOP/s)\end{tabular} & \begin{tabular}[c]{@{}c@{}}Parameter\\ learning\\ (MFLOP)\end{tabular} \\ \hline
\begin{tabular}[l]{@{}c@{}}TX leakage signal\\ w/o RF cancellation\end{tabular}             & -36                                                                    & 0                                                                               & 0                                                                      \\ \hline
\begin{tabular}[l]{@{}c@{}}TX leakage signal\\ after linear RF\\cancellation\end{tabular} & -62                                                                    & 7                                                                              & 34                                                                    \\ \hline
\begin{tabular}[l]{@{}c@{}}TX leakage signal\\ after nonlinear RF\\cancellation\end{tabular} & -85                                                                    & 31                                                                              & 135                                                                    \\ \bottomrule
\end{tabular}
\end{table}

\begin{figure}[!t]
\begin{minipage}[b]{1.0\linewidth}\centering
	\includegraphics[scale = 0.52]{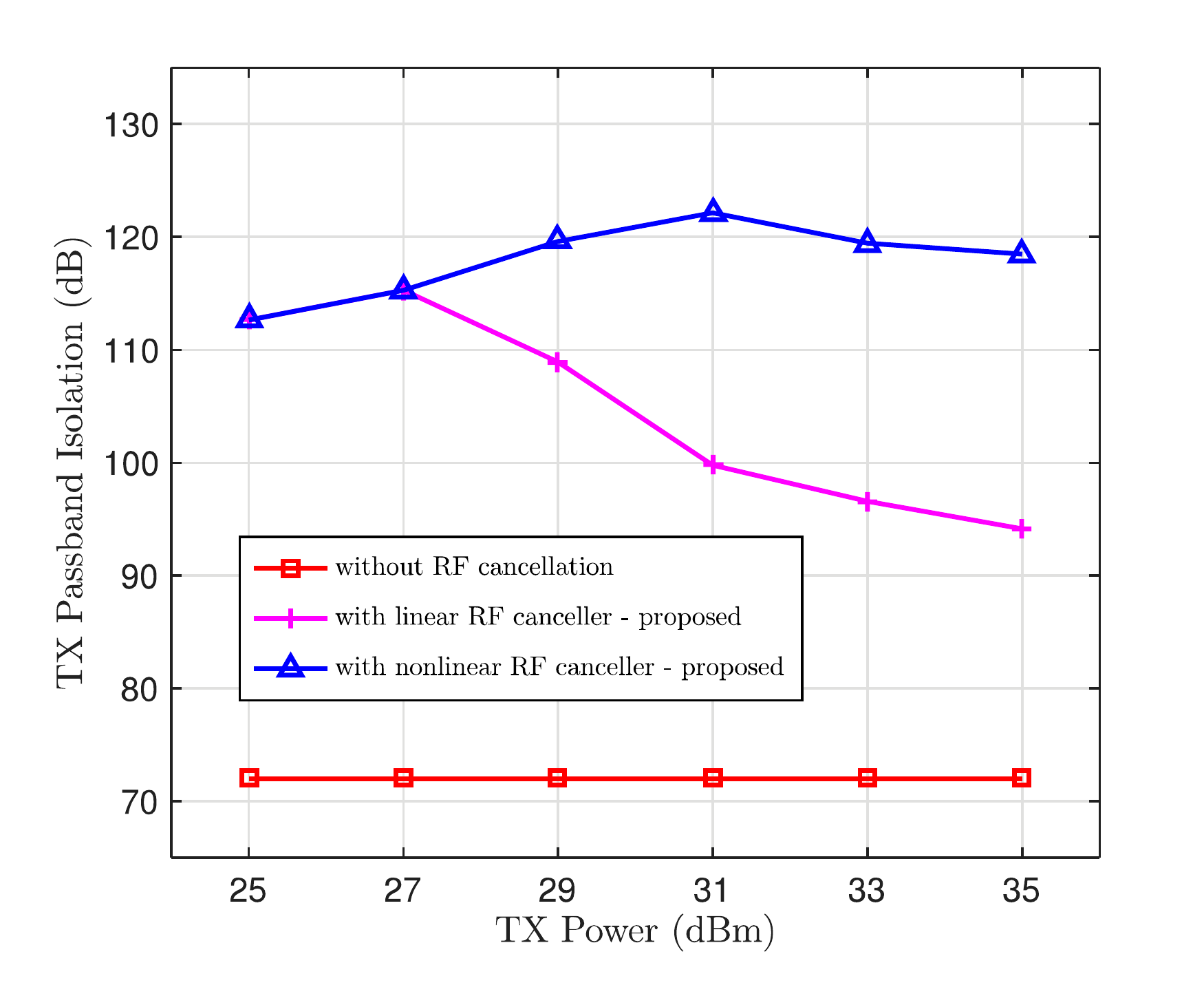}
\end{minipage}
\caption{{Comparison of the measured TX passband isolation against different TX powers with and without the proposed RF cancellation solution for the duplexer-based FDD case at $2.14$ GHz. TX signal bandwidth is $20$ MHz.}}\label{Meas_isolation}
\end{figure}

Fig.~\ref{Meas_isolation} shows the total achievable isolation after the proposed active RF cancellation with respect to different transmit power levels, evaluated using a $20$ MHz transmit signal bandwidth. The total isolation here refers to the sum of the duplexer average isolation and the RF cancellation gain. For reference, the average isolation of the commercial BS duplexer module used in the measurements is also plotted in the figure. As the curves in the figure show, the proposed cancellation solution can provide as much as $50$ dB of additional isolation, and the nonlinear canceler improves the cancellation gain by up to $25$ dB  at higher transmit power levels when compared to the linear canceler. Another observation is that at low transmit power levels, both the linear and the nonlinear RF cancellers have very similar cancellation performance. This is natural because the PA is still being operated in its linear region at lower transmit powers and the nonlinear distortion products are weak. Moreover, the cancellation gain is limited by the system noise floor at the lower transmit power levels.

\subsection{Circulator-Based Measurement Results}
\begin{figure}[!t]
\begin{minipage}[b]{1.0\linewidth}\centering
	\includegraphics[scale = 0.52]{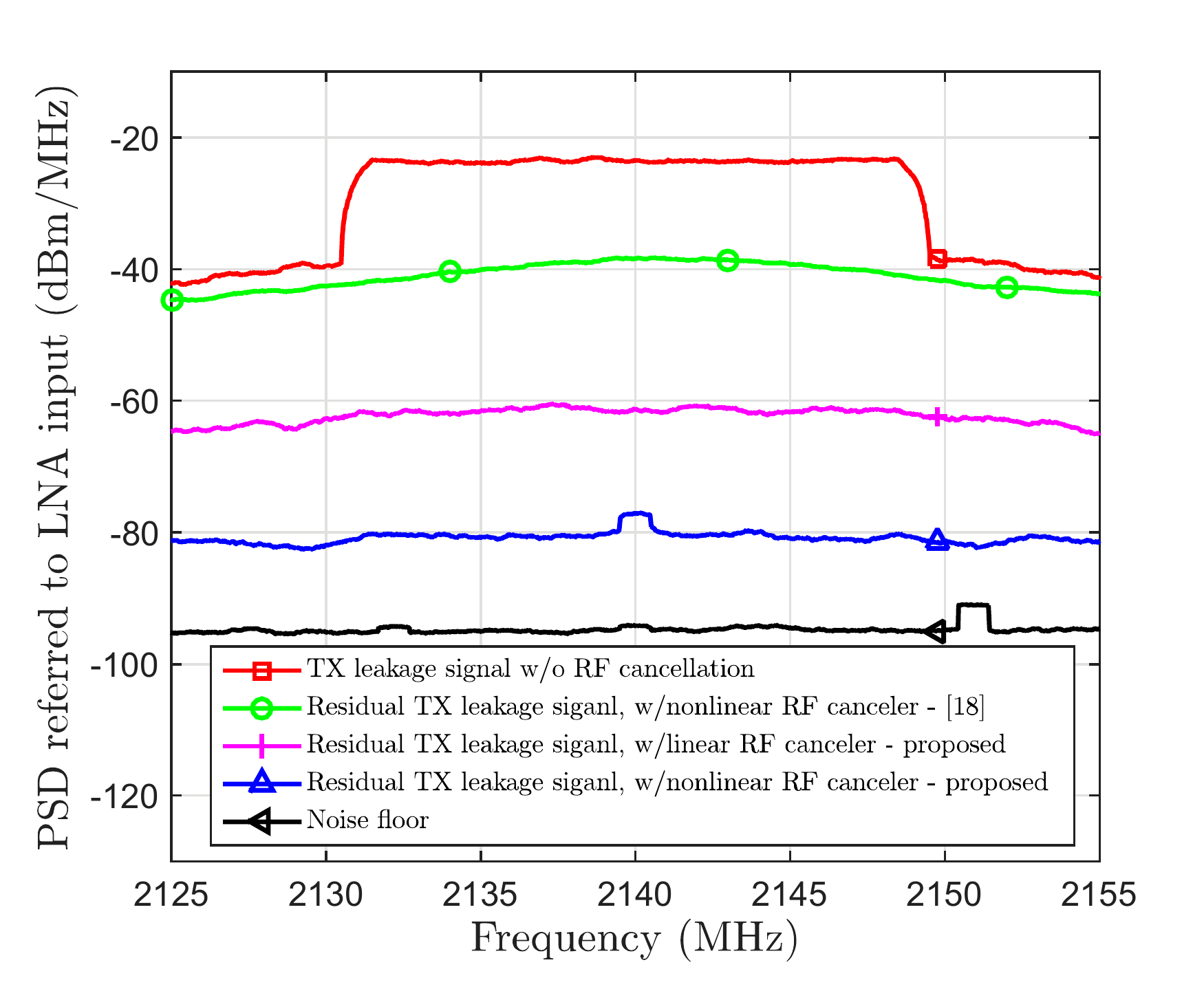}
\end{minipage}
\caption{{Measured signal spectra of the nonlinear TX leakage signal before and after different RF cancellers for an in-band FD transceiver example. TX signal bandwidth is $20$ MHz, TX power is $+30$ dBm, the total passive isolation against the transmit signal is $40$ dB, obtained through a circulator. LNA IIP3 is -7 dBm, while the TX center-frequency is $2.14$ GHz}}\label{Meas_Circulator}
\end{figure}

Next, we demonstrate and evaluate the capability of the proposed RF cancellation technique in an IBFD transceiver setting by adopting only a circulator as the passive isolation element, while tuning the TX and RX to the same center-frequency of 2.14 GHz. Compared to a duplexer which has a frequency-selective response and operates over fixed TX and RX frequency bands, a circulator has typically a milder frequency selectivity and operates over a wider frequency range, providing elementary passive isolation in IBFD communications systems where a shared TX/RX antenna is used. The adopted circulator provides $40$ dB isolation, the transmit power is $+30$ dBm, and correspondingly the average power of the transmit leakage signal at LNA input is thus $-10$ dBm. In the RX chain, we now utilize a highly nonlinear LNA (model no. HD24089) which has $22$ dB gain and IIP3 of -7 dBm. This experimental setting will enable us to evaluate the performance of the proposed cancellation technique under severe TX and RX chain nonlinearities. The nonlinear canceller order is now set to $P = 9$, whereas the cancellation filter length is $11$ taps for each basis function. The cancellation performance of different nonlinear RF cancellers with a $20$ MHz transmit signal bandwidth is depicted in Fig.~\ref{Meas_Circulator}, and the performance measures are reported in Table~\ref{comp_IBFD}. In this IBFD case, the TX leakage is approximately $70$ dB above the receiver effective thermal noise floor. Notice from the figure that the observable leakage signal, without RF cancellation, has significantly high distortion due to coexisting PA and LNA nonlinearities. The performance of the RF canceller proposed in \cite{Adnan_GlobalSIP} is limited by the LNA-induced distortion, and is able to suppress the TX leakage by only $17$ dB because the LNA nonlinearity is heavily limiting the parameter estimation performance. On the other hand, the proposed cancellation and closed-loop learning technique demonstrates that even a linear canceller can achieve up to $23$ dB of TX leakage suppression, while the nonlinear canceller gives then close to $54$ dB of RF cancellation, thus pushing down the self-interference to within $15$ dB of the system noise floor. The remaining residual self-interference can then be further suppressed by the existing purely digital SI cancellers, such as the one reported in \cite{Korpi2}, except for the purely random TX noise. Notice also that the system noise floor in Fig.~\ref{Meas_Circulator} is higher compared to the FDD measurement results reported in previous subsection due to significantly strong TX leakage signal. In general, the TX noise present in the residual self-interference cannot be cancelled by the digital canceller and, as discussed above, its impact can only be minimized through careful RF design, or alternatively, through digital cancellation approaches where the reference signal is taken from the PA output \cite{Choi3}, \cite{Huusari}, \cite{Su}, \cite{Zhou}. The measurement results of this experiment are also well in line with the theoretical framework and the proposed closed-loop parameter learning algorithm developments, indicating that the proposed cancellation filter parameter learning approach is indeed immune to the LNA-induced distortion. Furthermore, to the best of the authors knowledge, the obtained 54 dBs of measured RF cancellation represents state-of-the-art in active RF cancellation literature.

\begin{table}[]
\small
\centering
\caption{Comparison of the nonlinear TX leakage signal suppression with proposed active RF cancellation technique in a circulator-based IBFD transceiver experiment. TX power is $+30$ $\mathrm{dBm}$, circulator isolation is $40$ $\mathrm{dB}$, and the transmit signal bandwidth is $20$ $\mathrm{MHz}$. Also the involved processing complexities are shown.}
\label{comp_IBFD}
\begin{tabular}{l|c|c|c}
\toprule
\multirow{2}{*}{}                                                                           & \multirow{2}{*}{\begin{tabular}[c]{@{}c@{}}\\ Power\\ (dBm)\end{tabular}} & \multicolumn{2}{c}{Complexity}                                                                                                                          \\ \cline{3-4} 
                                                                                            &                                                                        & \begin{tabular}[c]{@{}c@{}}Interference\\ regeneration\\ (GFLOP/s)\end{tabular} & \begin{tabular}[c]{@{}c@{}}Parameter\\ learning\\ (MFLOP)\end{tabular} \\ \hline
\begin{tabular}[l]{@{}c@{}}TX leakage signal\\ w/o RF cancellation\end{tabular}             & -10                                                                    & 0                                                                               & 0                                                                      \\ \hline
\begin{tabular}[l]{@{}c@{}}TX leakage signal\\ after linear RF\\cancellation\end{tabular} & -45.7                                                                    & 6                                                                              & 29                                                                    \\ \hline
\begin{tabular}[l]{@{}c@{}}TX leakage signal\\ after nonlinear RF\\cancellation\end{tabular} & -64.3                                                                    & 40                                                                              & 143                                                                    \\ \bottomrule
\end{tabular}
\end{table}

\begin{figure}[!t]
\begin{minipage}[b]{1.0\linewidth}\centering
	\includegraphics[scale = 0.52]{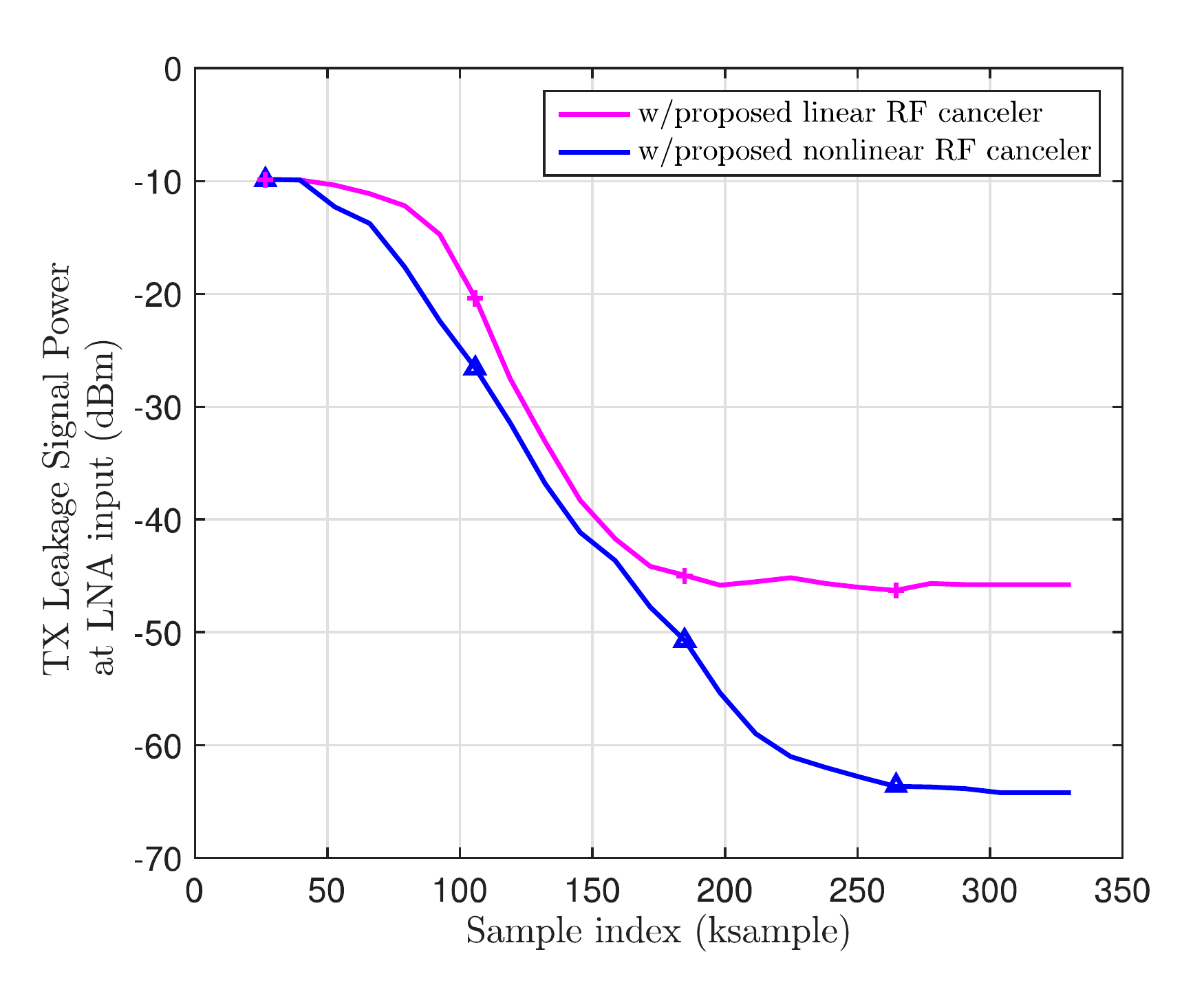}
\end{minipage}
\caption{{Measured instantaneous power of the TX leakage signal at the LNA input after active RF cancellation with respect to the sample index in an IBFD transceiver example. TX signal bandwidth is $20$ MHz, TX power is $+30$ dBm, the circulator passive isolation against the transmit signal is $40$ dB. LNA IIP3 is -7 dBm, while the TX center-frequency is $2.14$ GHz}}\label{Meas_PwrvsTime}
\end{figure}

Finally, the convergence behavior of the proposed RF cancellation technique is illustrated in Fig.~\ref{Meas_PwrvsTime} by plotting the measured instantaneous power of the nonlinear TX leakage signal at the LNA input against the transmit data sample index. The results here indicate that, for the utilized estimation block of $13.000$ samples, the closed-loop learning system requires approximately $20$ iterations to converge to steady-state coefficients, or alternatively about $4$ msec in a real-time processing system in this example case. The fast convergence and stable operation can be partly attributed to the orthogonalization of the basis functions as explained in Section IV-A.

In general, the presented active RF cancellation results are clearly state-of-the-art, showing that the total achievable isolation, by combining the passive isolation and RF cancellation, is more than $90$ dB, which leaves only some $15$ dB of SI cancellation for the digital canceller. The measurement results also highlight the clear advantage of nonlinear processing and the proposed closed-loop parameter learning approach for estimating the cancellation filter coefficients under a practical nonlinear LNA in the loop. 

\section{Conclusion}
Transmitter-induced self-interference is a major challenge in simultaneous transmit-receive systems, and obtaining sufficient TX-RX isolation is crucial to enable the proper operation of the receiver. In this paper, we proposed a nonlinear active RF cancellation technique for TX leakage suppression that can complement the elementary passive isolation, thereby substantially improving TX-RX isolation and enabling flexible and efficient spectrum utilization. In the proposed scheme, we first regenerate a complex baseband estimate of the true RF TX leakage signal in the transceiver digital front-end, through nonlinear filtering of the known transmit data. The actual RF cancellation signal is then generated through an auxiliary transmit path, and added to the received signal at the RX LNA input, such that the nonlinear TX leakage is suppressed. Furthermore, a novel closed-loop decorrelation-based algorithm was presented to estimate the cancellation filter coefficients in an efficient manner. Unlike other works in the existing literature, the proposed nonlinear canceller and the closed-loop parameter learning system were shown to tolerate the LNA-induced nonlinear distortion of the TX leakage signal, and thus provide enhanced cancellation performance. Also the computational complexity of the proposed solution was addressed and shown to be feasible for today's base-station processing units. We evaluated the performance of the proposed cancellation scheme with comprehensive RF measurements, adopting LTE-Advanced BS transceiver components and incorporating both FDD and IBFD measurement scenarios. The measured results indicate that the proposed nonlinear canceller can achieve beyond $50$ dBs of self-interference and TX leakage suppression, representing state-of-the-art. Hence, the proposed scheme can enable the adoption of very simple and compact duplexers or other potential low passive isolation circuits in the future radio devices. Such efficient self-interference cancellation schemes can also be seen as one potential technique to support flexible spectrum allocation and utilization in future $5$G radio networks.

\end{document}